\documentstyle[prl,floats,twocolumn,aps,epsfig,psfrag]{revtex}
\tighten
\begin{document}
\draft
\twocolumn[\hsize\textwidth\columnwidth\hsize\csname @twocolumnfalse\endcsname
\title{A toy model for slowly growing wormholes as effective topology 
changes}

\author{Samuel L.~Braunstein}
\address{Universit\"at Ulm, Abteilung Quantenphysik, 89069 Ulm, Germany}
\address{\cite{CurrAd}SEECS, University of Wales, Bangor, Gwynedd LL57 1UT, UK}

\date{\today}
\maketitle

\begin{abstract}
We present a toy model for growing wormholes as a model of effective
low-energy topology changes. We study the propagation of quantum 
fields on a $1+1$ spacetime analogous to the trouser-leg topology 
change. A low-energy effective topology change is produced by a
physical model which corresponds to a barrier smoothly changing
the tunneling probability between two spatial regions.
\end{abstract}
\vspace{3ex}
]

In 1957 Wheeler argued that the topology of space must fluctuate at what
he called the ``Planck scale'' of distances \cite{Wheeler}. He concluded 
that in a linearized 
theory of gravity the Riemann tensor would exhibit fluctuations violent 
enough to practically pinch off portions of space or produce multiply 
connected ``wormholes.'' Even though the scales involved meant that the 
linearized theory would fail, Wheeler suggested that these phenomena 
would, nonetheless, occur.

Interest in the possibility of wormholes was greatly increased by 
the Morris-Thorne wormhole solution \cite{MorrisThorne} of classical 
general relativity. They found that if a wormhole can be formed or 
found by accident then it can in principle be stretched to allow 
`faster-than-light' travel or time-travel \cite{Morris2}. Indeed, 
one of the most bizarre possibilities of wormholes would be to escape 
from the gravitational pull of a black-hole from well within its 
Schwarzschild radius \cite{FrolovNovikov}!

The possibility of creating wormhole structures faces severe problems: 
Classically, the weak energy condition must be violated, both to supply 
suitable matter for the construction of a Morris-Thorne wormhole and to 
circumvent classical singularity theorems \cite{Geroch,Tipler}. However,
quantum mechanics allows for violation of the weak energy condition
\cite{Casimir,Braunstein}. Unfortunately, quantum fields propagating
on a time varying topology introduce new problems: Anderson and DeWitt 
\cite{AndersonDeWitt} performed the first model calculations with a 
quantum field propagating on a classical spacetime manifold which 
instantaneously changed its topological connectedness. They showed
that this sort of ideal topology change leads to the production
of infinitely bright flashes of energy \cite{Manogue,HarrisDray}.

\begin{figure}[thb]
\epsfxsize=3.2in
\epsfbox[85 350 508 435]{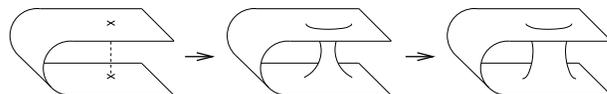}
\caption{A microscopic wormhole is gradually stretched. Initially, the 
low-energy effective theory is spacetime without a wormhole. However, as 
the wormhole is stretched the tunneling probability for successively 
lower energy modes approaches one---leaving spacetime with a wormhole.}
\label{THEfig1}
\end{figure}

How might we evade an instantaneous topological change? We argue 
that an effective low-energy theory can behave exactly like a gradual
topology change. The fundamental idea is that the propagating fields
should see the change occur only at the amplitude level: Quantum fields 
propagating on the spacetime experience a gradually increasing tunneling 
probability from the original to the new `topological' configuration. For 
wormhole `creation' this could be equivalent to slowly stretching a 
microscopic wormhole to a large size. See Fig.~\ref{THEfig1}. For such 
a gradual stretching the low-energy dynamics of fields on this background 
spacetime would evolve smoothly from those on a spacetime without a 
wormhole to those with one. Initially, only modes with energies above 
the Planck-scale can tunnel between the upper and lower sheets, however, 
this path becomes more and more accessible to low-energy modes as the 
wormhole is gradually enlarged. The growing wormhole is acting as a
time-varying waveguide. This procedure delays the problem 
discovered by Anderson and DeWitt until the Planck-scale where it is 
no longer applicable. As far as the low-energy effective theory is 
concerned this would be an effective topology change, whether or not it 
was fundamentally so. The important point is that the fields see smoothly
changing tunneling probabilities instead of a discontinuous change
in the underlying topology. In fact, any mechanism which yields this
behavior might mediate a low-energy effective topology change, in
this sense our proposal is more general than merely slowly growing a
classical wormhole.

\begin{figure}[thb]
\epsfxsize=3.0in
\epsfbox[-90 -15 180 120]{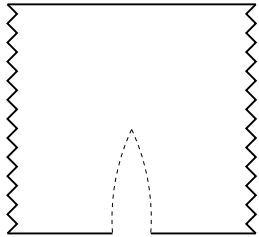}
\caption{Spacetime diagram around the `crotch' (dashed line) of our 
toy `wormhole.' Time is running up. As the barrier height $2 V(t)$ is 
gradually lowered from being huge to being zero the disconnected 
half-spaces slowly reconnect.}
\label{THEfig2}
\end{figure}

Calculations along the lines of those described in Fig.~\ref{THEfig1} 
involving an actual growing wormhole are too hard, so instead we 
consider a $1+1$ dimensional toy model with Lagrangian density as
\begin{equation}
{\cal L}={1\over 2}\left[
\left(\partial\phi\over\partial t\right)^2 -
\left(\partial\phi\over\partial x\right)^2
-2 V(t) \delta(x) \phi^2\right] \label{Lag} \;,
\end{equation}
where $2V(t)$ represents the `height' of the tunneling barrier between 
the left and right half-spaces. As the barrier height is slowly 
shrunk from being huge to being zero the initially effectively 
disconnected half-spaces reconnect. In some sense this is a 
good model for the physics we are interested in for a growing wormhole 
in that the field modes see a gradually increasing tunneling probability 
between the half-spaces, and that only the high frequency modes can 
effectively tunnel the barrier which otherwise strongly reflects the low 
frequency modes. However, in other respects this model is not so similar 
to the wormhole scenario, and instead resembles a sort of opened-string 
pants diagram which we might draw as Fig.~\ref{THEfig2}. The crotch has
been drawn as sort of hazy to remind us that the geometric optics limit
is failing and wavepackets are dispersively reflected and transmitted
during the time-varying dynamics of the barrier. Finally, we note that
our toy model is closely analogous to the trousers spacetime originally 
considered by Anderson and DeWitt in their study of the behavior of 
quantum fields propagating on a time-varying topology \cite{AndersonDeWitt}.

The equations of motion are the wave-equation in the left and
right half-spaces and an equation for the `boundary' conditions
that must be satisfied at $x=0$
\begin{equation}
\partial_x \phi(0^+,t)-\partial_x \phi(0^-,t) = 2 V(t) \phi(0,t)\;. \label{BCs}
\end{equation}
We assume continuity of the field across $x=0$ and use so-called 
light-cone coordinates $u=t-x$ and $v=t+x$ to write down the
general solution for $\phi$ as
\begin{equation}
\phi(x,t)=\left\{\begin{array}{ll}
    U_{\rm in}(u)+V_{\rm out}(v)\;,~~& x<0 \;, \\
    U_{\rm out}(u)+V_{\rm in}(v)\;,& x\ge 0 \;.
  \end{array} \right. \label{GenForm}
\end{equation}
Eliminating $U_{\rm out}'(u)$ and $V_{\rm out}'(v)$ from the
boundary equation we find
\begin{equation}
\partial_t \phi(0,t) + V(t) \phi(0,t) = U_{\rm in}'(t)+V_{\rm in}'(t) 
\label{BCeq} \;.
\end{equation}
The general solution at $x=0$ is
\begin{eqnarray}
\phi(0,t) &=& \phi(0,t_0) \,e^{-\int_{t_0}^{t} dt'\,V(t')}\nonumber \\
&&+ \int_{t_0}^t dt' e^{-\int_{t'}^{t} dt''\,V(t'')}
\left[U_{\rm in}'(t)+V_{\rm in}'(t)\right] \label{GenSol} \;.
\end{eqnarray}
The full solution throughout space is then given by
\begin{equation}
\phi(x,t)=\left\{\begin{array}{ll}
    U_{\rm in}(u)+\phi(0,v)-U_{\rm in}(v)\;,~~& x<0 \;, \\
    \phi(0,u)-V_{\rm in}(u)+V_{\rm in}(v)\;,& x\ge 0\;.
  \end{array} \right. \label{PropSol}
\end{equation}

Let us consider an adiabatic expansion of $\phi(0,t)$ for an
incoming harmonic wave: e.g., $U_{\rm in}(t)=e^{-i\omega t}$
and $V_{\rm in}(t)=0$. To first order in the time-derivative of the 
barrier height we find:
\begin{eqnarray}
\phi(0,t)&\simeq&{i\omega \over i\omega -V(t)}\, e^{-i\omega t} \left(
1+ {V'(t)\over[i\omega -V(t)]^2}\right)\nonumber \\
&& - {i \omega V'(t_0)e^{-i\omega t_0}\over[i\omega -V(t_0)]^3} 
\label{adiabatic} \;,
\end{eqnarray}
which may be checked by direct substitution into the differential
equation~(\ref{BCeq}). For the model barrier time-dependence 
\begin{equation}
V(t) = {V_0 \,e^{-\omega_0 t} + V_1\, e^{\omega_0 t}\over
e^{-\omega_0 t}+e^{\omega_0 t}} \label{modelTD} \;,
\end{equation}
we find that the first-order derivative terms in~(\ref{adiabatic}) are 
small for all $t$ and $\omega$ when
\begin{equation}
\omega_0 \ll {\min (V_0, V_1) \over |V_1-V_0|} \label{adiab} \;,
\end{equation}
i.e., for a sufficiently slow change in barrier height we have obtained
a uniformly accurate approximation to the general solution of
Eq.~(\ref{GenSol}).

\begin{figure}[thb]
\begin{psfrags}
\psfrag{time}[l]{\Large$t$}
\psfrag{space}[r]{\Large$~~~x$}
\psfrag{b1}[c]{\Large$e^{-i\omega u}$}
\psfrag{c1}[r]{\Large$\rho\, e^{-i\omega v}$}
\psfrag{c2}[l]{\Large$\tau e^{-i\omega u}$}
\epsfxsize=3.4in
\epsfbox[0 30 244 150]{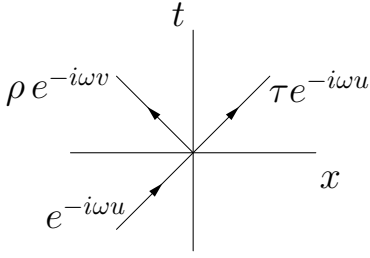}
\end{psfrags}
\caption{A `spacetime' diagram of the right-headed wave $e^{-i\omega u}$
with a static barrier: It is partially reflected off the barrier
into a left-headed wave $\rho\, e^{-i\omega v}$ and partially
transmitted into $\tau e^{-i\omega u}$.}
\label{bs1}
\end{figure}

Our Lagrangian density~(\ref{Lag}) has an associated time-independent 
inner product
\begin{equation}
(u_1, u_2) \equiv {i\over 2}\int_{-\infty}^\infty dx\, 
(\overline{u}_1\,\partial_t u_2
-\partial_t\overline{u}_1 u_2)\label{IP} \;,
\end{equation}
where the overline represents complex conjugation. Using this 
and the result that
$$
\int_0^\infty dx\, e^{\pm i\omega x}
= \pi\delta(\omega) \pm i\,{\cal P} {1\over \omega} \;,
$$
where ${\cal P}$ represents the Cauchy principle value function,
we may determine the in-coming traveling wave mode solutions
for a {\it static\/} barrier. The solutions are
\begin{eqnarray}
u_\omega^{\rm in}&=&{1\over\sqrt{2\pi\omega}}\left\{
  \begin{array}{ll}
    e^{-i\omega u}+ \rho\, e^{-i\omega v}\;,~~& x<0 \;, \\
    \tau e^{-i\omega u}\;,& x\ge 0 \;, \\
  \end{array} \right. \nonumber \\
v_\omega^{\rm in}&=&{1\over\sqrt{2\pi\omega}}\left\{
  \begin{array}{ll}
    \tau e^{-i\omega v}\;,& x< 0 \;, \\
    e^{-i\omega v}+ \rho\, e^{-i\omega u}\;,~~& x\ge 0 \;, \\
  \end{array} \right. \label{InStatic}
\end{eqnarray}
so the static barrier has reflection and transmission coefficients
\begin{equation}
\rho=\tau-1 ~~~{\rm and}~~~
\tau={i\omega\over i\omega-V} \label{RTCoeffs} \;,
\end{equation}
respectively (thus, $|\rho|^2 +|\tau|^2=1$ so energy is conserved
in this case). The right-headed
incoming mode $u_\omega^{\rm in}$ is illustrated in Fig.~\ref{bs1}.
These modes form a complete orthonormal set according to the inner
product~(\ref{IP}) with
\begin{eqnarray}
(u_\omega^{\rm in}, u_{\omega'}^{\rm in}) &=&
(v_\omega^{\rm in}, v_{\omega'}^{\rm in}) = \delta (\omega-\omega')  \;,
\nonumber \\
(u_\omega^{\rm in}, v_{\omega'}^{\rm in}) &=&
(u_\omega^{\rm in}, \overline{u}_{\omega'}^{\rm in}) =
(u_\omega^{\rm in}, \overline{v}_{\omega'}^{\rm in}) = 0 \;.
\end{eqnarray}

Time-reversal invariance implies that the outgoing solutions have the form
\begin{eqnarray}
u_\omega^{\rm out}&=&{1\over\sqrt{2\pi\omega}}\left\{
  \begin{array}{ll}
    \overline{\tau} e^{-i\omega u}\;,& x<0 \;, \\
    e^{-i\omega u}+ \overline{\rho}\, e^{-i\omega v}\;,~~& x\ge 0 \;, \\
  \end{array} \right. \nonumber \\
v_\omega^{\rm out}&=&{1\over\sqrt{2\pi\omega}}\left\{
  \begin{array}{ll}
    e^{-i\omega v}+ \overline{\rho}\, e^{-i\omega u}\;,~~& x<0 \;, \\
     \overline{\tau} e^{-i\omega v}\;,& x\ge 0 \;. \\
  \end{array} \right. \label{Uout}
\end{eqnarray}

Canonical quantization of the field proceeds by enforcing the standard
equal-time commutation relations with conjugate momentum
$\hat\pi(x,t)\equiv\partial\hat\phi(x,t)/\partial t$ via
\begin{equation}
[\hat\phi(x,t),\hat\pi(x',t)]=i\,\delta(x-x') \;,
\end{equation}
where for convenience we choose Planck's constant as unity.
This language may be translated to that of mode annihilation and
creation operators by decomposing the field into any complete orthonormal
set of modes with respect to the inner product~(\ref{IP}), for instance
\begin{equation}
\FL
\hat\phi(x,t)={1\over\sqrt{2}}\int_{0}^\infty 
d\omega \left[u_\omega(x,t)\,\hat a_\omega
+ v_\omega(x,t)\,\hat b_\omega + {\rm h.c.}\right] \;,
\end{equation}
with $u_\omega$ and $v_\omega$ denoting full solutions to the 
classical equations of motion for right- and left-headed 
traveling-wave modes, respectively. By virtue of the 
equal-time commutation relations for $\hat\phi$ and $\hat\pi$ 
the annihilation and creation operators satisfy the canonical 
commutation relations
\begin{equation}
[\hat a_\omega,\hat a_{\omega'}^\dagger]=
[\hat b_\omega,\hat b_{\omega'}^\dagger]= \delta(\omega-\omega') \;,
~~~~[\hat a_\omega,\hat b_{\omega'}]=0 \;.
\end{equation}
This completes the canonical quantization for our one-dimensional field.

The Bogoliubov transformations may be derived in the following way:
Using the time-evolved incoming modes with $V=V(t)$ and
the static outgoing modes [Eq.~(\ref{Uout})] with $V$
replaced by the final value $V_1$ we obtain an expression for 
the annihilation operator for the outgoing modes as
\begin{eqnarray}
\hat a_\omega^{\rm out} &=&
\int_{0}^\infty d\omega'
\left[ \left.(u_\omega^{\rm out}, u_{\omega'}^{\rm in} )\right|_{t_1}
\hat a_{\omega'}^{\rm in} +
\left.(u_\omega^{\rm out}, v_{\omega'}^{\rm in} )\right|_{t_1}
\hat b_{\omega'}^{\rm in} \right. \nonumber \\
&&\phantom{\int}\left.+\left.(u_\omega^{\rm out},
\overline{u}_{\omega'}^{\rm in} )\right|_{t_1}
\hat a_{\omega'}^{\rm in\dagger} +
\left.(u_\omega^{\rm out}, \overline{v}_{\omega'}^{\rm in} )\right|_{t_1}
\hat b_{\omega'}^{\rm in\dagger} \right] \;.\!
\end{eqnarray}
In fact, it is sufficient for the purposes of determining the
number of quanta generated to calculate the latter two terms
alone. We may determine $\hat b_\omega^{\rm out}$ by symmetry.

Consider the time-evolved mode $u_\omega^{\rm in}$: It's Bogoliubov 
coefficient with the static mode $u_{\omega'}^{\rm out}$ is given by
\begin{eqnarray}
\left(\overline{u}_{\omega}^{{\rm in}(t)}, u_{\omega'}^{\rm out}\right) &=&
{1\over2\pi\sqrt{\omega\omega'}}
\Bigl( (V_1-V_0) {\tau_0 \overline{\tau}'_1 \over\omega+\omega'}
e^{-i(\omega+\omega')t_1} \nonumber \\
&&~~\,-i\overline{\tau}'_1 e^{-i\omega' t_1}
\left[ \phi(0,t_1) -\phi_0(0,t_1)\right] \label{IPexp} \\
&&~~\,+ \omega'\!\!\int_{t_0}^{t_1}\!\! dt'\,e^{-i\omega' t'}
\left[\phi(0,t') -\phi_0(0,t')\right] \Bigr)\nonumber \,,
\end{eqnarray}
where $\phi_0(0,t)\equiv\tau_0 e^{-i\omega t}$ and subscripts $j=0,\,1$
refer to using $V=V_j$. The Bogoliubov coefficient for the 
time-evolved mode $v_\omega^{\rm in}$ with static 
$u_{\omega'}^{\rm out}$ is identical.

How many quanta are produced and with what spectrum? The power spectrum 
of right-going quanta is given by
\begin{eqnarray}
\langle \hat n_\omega(R)\rangle &=&
\left\langle \hat a_\omega^{\rm out\, \dagger}\,\hat a_\omega^{\rm out}
\right\rangle \nonumber \\
&=&2 \int_0^\infty d\omega'\left.
\left| \left(u_\omega^{\rm out},\,\overline{u}_{\omega'}^{\rm in}\right)
\right|^2 \right|_{t=t_1} \label{PS} \;.
\end{eqnarray}
By symmetry the left- and right-going quanta have the same power spectra.

\begin{figure}
\epsfbox[5 415 260 530]{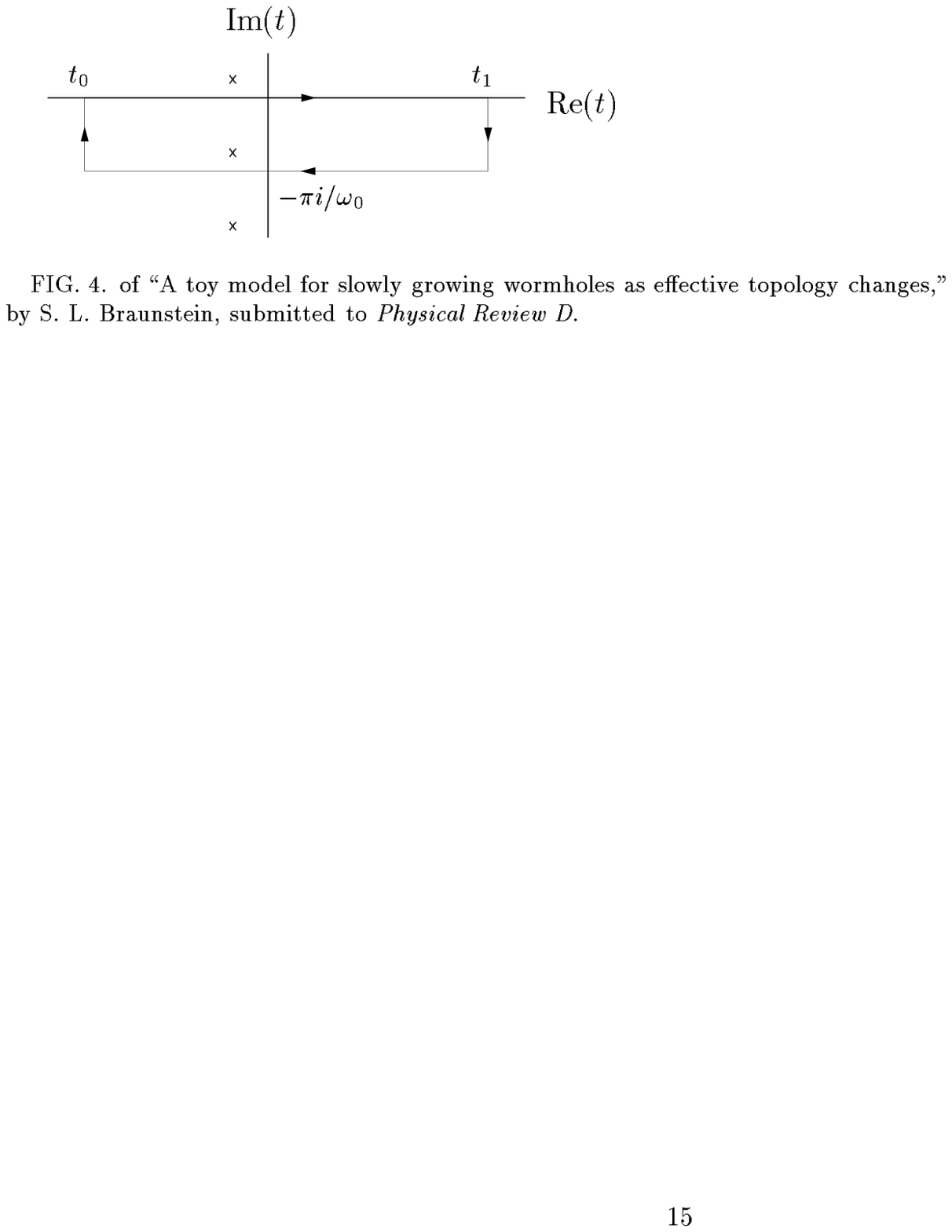}
\caption{Contour used to perform integral~(\protect\ref{contint}).
The crosses denote poles.}
\label{contour}
\end{figure}

Consider the model time-dependence for the barrier given by 
Eq.~(\ref{modelTD}). Taking the lowest order adiabatic expansion as
\begin{equation}
\phi(0,t) \simeq {i \omega\over i\omega -V(t)} \,e^{-i\omega t} \;,
\end{equation}
and the limit $t_0\rightarrow -\infty$ we have
\begin{eqnarray}
\left(\overline{u}_{\omega}^{{\rm in}(t)}, u_{\omega'}^{\rm out}\right) 
&\simeq& {\sqrt{\omega\omega'}(V_1-V_0)\over2\pi(i\omega-V_0)}
\Biggl( {e^{-i(\omega+\omega')t_1} \over(\omega+\omega')(i\omega-V_1)} 
\nonumber \\
&&+ i\!\!\int_{-\infty}^{t_1}\!\!\!\! dt'\,{e^{-i(\omega+\omega') t'}
\over(i\omega-V_0)e^{-2\omega_0 t'}+(i\omega-V_1)}\Biggr)\nonumber . \\
\end{eqnarray}
This integral may be performed with a suitablely chosen contour
(see Fig.~\ref{contour}) yielding for $t_1$ asymptotically large:
\begin{eqnarray}
&&i\int_{-\infty}^{t_1} dt\,{e^{-i\Omega t}
\over A_0\,e^{-2\omega_0 t}+A_1} \label{contint} \\
&=&{-e^{-i\Omega t_1}\over \Omega A_1} + 
{\pi\over 2\omega_0 A_1 \sinh (\pi\Omega /2 \omega_0)}
\left(A_1\over A_0\right)^{i\Omega\over 2\omega_0} \nonumber \;.
\end{eqnarray}
The inner-product then reduces to
\begin{eqnarray}
\left(\overline{u}_{\omega}^{{\rm in}(t)}, u_{\omega'}^{\rm out}\right)
&\simeq& {\sqrt{\omega\omega'}(V_1-V_0)\over 4\omega_0(i\omega-V_0)
(i\omega-V_1)} \\
&&\times{1\over\sinh [\pi(\omega+\omega') /2 \omega_0]}\!
\left(i\omega-V_1\over i\omega-V_0\right)^{\!i(\omega+\omega')
\over 2\omega_0} \nonumber
\end{eqnarray}
which as expected is independent of $t_1$.

The overlap between the distant-past incoming and distant-future
outgoing field modes is then
\begin{eqnarray}
&&\left|\left(u_{\omega}^{\rm out}, 
\overline{u}_{\omega'}^{{\rm in}(t)}\right) \right|^2 
\simeq {\omega\omega'(V_1-V_0)^2\over 16\omega_0^2(\omega'^2+V_0^2)
(\omega'^2+V_1^2)} \nonumber \\
&&~~\,\times{\exp\left\{{\omega+\omega'\over\omega_0}\left[
{\rm Arctan}(\omega'/V_1)-{\rm Arctan}(\omega'/V_0)\right]\right\}
\over\sinh^2 [\pi(\omega+\omega') /2 \omega_0]} \;.
\end{eqnarray}
The integral of this expression yields the  power spectrum~(\ref{PS}) 
and may be well approximated by noting that the only significant 
contributions come from $\omega'\ll V_0,\, V_1$:
\begin{eqnarray}
\langle\hat n_\omega(R)\rangle &\simeq& - {\omega(V_1-V_0)^2
\over 2\pi^2 V_0^2 V_1^2}\, \ln( 1 - e^{-\pi\omega/\omega_0}) \\
&\simeq& {(V_1-V_0)^2 \over 2\pi^2 V_0^2 V_1^2} \times \left\{
\begin{array}{ll}
-\omega\,\ln(\pi\omega/\omega_0) \;, ~~& \omega\ll\omega_0 \;, \nonumber \\
\omega \,e^{-\pi\omega/\omega_0} \;, & \omega \gg\omega_0\;. \nonumber
\end{array} \right.
\end{eqnarray}

The total energy generated is (recalling $\hbar=1$)
\begin{eqnarray}
{\cal E} &=& \int_0^\infty d\omega\,\omega [\langle\hat n_\omega(L)\rangle+
\langle\hat n_\omega(R)\rangle] \nonumber \\
&\simeq& -{\omega_0^3(V_1-V_0)^2 \over \pi^5 V_0^2 V_1^2}
\int_0^\infty dx\, x^2 \ln(1-e^{-x}) \nonumber \\
&=& {\omega_0^3(V_1-V_0)^2 \over 45 \pi V_0^2 V_1^2} 
\ll (\hbar)\,\omega_0 \;,
\end{eqnarray}
so the adiabatic response is very weak even as $V_0\rightarrow\infty$
and $V_1\rightarrow 0$. We note that both of these limits can only be 
taken within our calculations as a double limit with $\omega_0\rightarrow0$
thus ensuring the adiabaticity condition~(\ref{adiab}) continues to 
be satisfied.

We have studied a relativistic quantum field theory with tunneling
through a time-varying barrier. The smoothly changing tunneling
probability leads to an open-string pants spacetime with a
hazy `crotch.' Our toy model calculations suggest that slowly
growing a microscopic wormhole should appear as an effective 
low-energy topology change. Finally, we have shown that quantum 
fields propagating on such an effective topology change need not 
have the non-renormalizable difficulties associated with instantaneous 
topology changes of the underlying manifold.

\vskip 0.2truein

SLB was funded by a Humboldt Fellowship and appreciated the 
hospitality of the Institute for Theoretical Physics under National 
Science Foundation Grant No.~PHY94-07194 and of the Center for 
Advanced Studies at the University of New Mexico under Office
of Naval Research Grant No.~N00014-93-1-0116. He thanks James Anglin
for encouragement and Bill Unruh for discussions.

\newpage

\end{document}